# Time Served

Evan Milton, Mark Austin, Dan McArthur
Fermilab, Batavia, United States of America

*Abstract*

The following document provides a comprehensive look into the past, present, and future of Timing at Fermilab. By including historical context with technical implementation, this document seeks to help the reader gain a broader understanding of the facility and the mechanisms that enable the operation of Fermilab's accelerators.

## BACKGROUND

The Tevatron Clock (TCLK) Timing system has been an integral part of Fermilab's operation since its inception in 1980. Originally developed to unify the 5 different clocking systems already present at the lab [1], it has provided a reliable mechanism for the complex transfer scenarios required to run the Tevatron through 2012. Still in use today, the protocol provides both General Machine Timing (GMT) for the complex and Beam Synchronous (BS) Timing for the Main Injector/Recycler machines. This proven platform is very much the work horse of America's *premier* High Energy Physics (HEP) laboratory, and the seed for other Timing systems actively in use nationwide.

Before the advent of modern Network Time Protocols (NTP), TCLK provided a method to synchronize the actions of thousands of distributed accelerator devices within 2ns of the beam itself [2]. This systems saw the Fermilab accelerator complex through the entirety of collider operations and into the Main Injector Era; signaling the transfers, ramps, resets, and extractions that ultimately delivered the Top and Bottom Quarks. Despite its 'outdated' reputation, machine experts and operators alike are in general agreement: TCLK does what it was put there to do. Simply put: "If it ain't broke, don't fix it" and for this reason, we treat Timing with the upmost respect in this document.

## PAST

The following section was taken from Bob Ducar and David Beechy's seminal paper "Time and Data Distribution Systems at the Fermilab Accelerator" c. 1983 [1]

> *The operation of the accelerator facility at Fermilab has been likened to the functioning of a vast multichannel predetermined timer. Hundreds of electronic devices must function in a time dependent fashion for the various accelerators to operate separately and together smoothly. As was the case for most previous accelerator facilities, specialized clock systems were developed at Fermilab to support the Linac, Booster, and Main Ring to orchestrate the operation of these machines. As the Main Ring become operational in 1972, separate clocks existed for the different accelerators - each utilizing different base band frequencies and encoding techniques. […] All of these clocks incorporated coded markers that signified the reset of an acceleration cycle or some other major Event within the cycle. A variety of techniques were used to encode these markers - most notably gaps, phase reversals, or both.*
>
> *The advent of planning and construction of the Tevatron presented an opportunity to review the existing clock facilities. A total of five different clock systems were implemented at the time - generally not compatible but magically connected together with a myriad of cables to operate in a coordinated fashion. This was clearly not a base on which to provide the stringent Timing requirements that the Tevatron demanded. Clocks do not disappear overnight. The designers followed the only reasonable course and began the specification of a sixth clock system - the Tevatron Clock.*

As stated by Ducar & Beechy, early accelerator clocks at Fermilab were comparatively primitive; accomplishing only one or two functions through a 'magically connected myriad of cables' that coordinated phase reversal and gap transfers between machines. The original architect of the system, Bob Ducar, acknowledges that this would be the '6th attempt at a standard' and to his great credit the protocol stuck. TCLK is presently the only Timing protocol in use at the lab today.

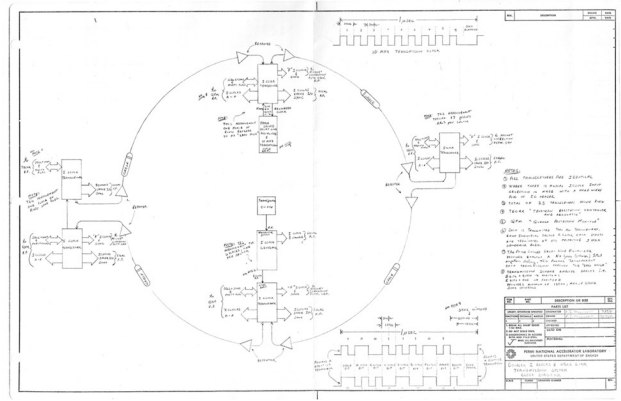

*Figure 1: Original "ICLK" Distribution for the Tevatron*

## HOW TIMING WORKS AT FERMILAB

Fermilab is a particle accelerator complex, and particle accelerators are complex. Starting with the resonant regulation infrastructure controlling Booster's bend fields (see the concepts rookie book [3]) the lab operates as a real-time state machine, with zero crossings from the Master Substation putting into action a complex series of events across campus. By tying a sequence of actions (timeline) directly to mains power, operators are able to tune the machine to near-optimal efficiency. This interaction with the 60Hz mains power is shown in an early form below.

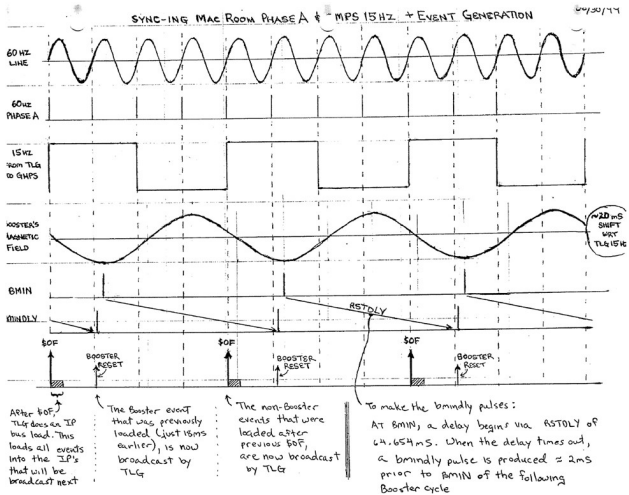

*Figure 2: Booster Timing at Fermilab. Triplett, c.1999*

## IMPLEMENTATION

TCLK utilizes a 'Modified Manchester' encoding that encapsulates both a carrier frequency and 8bit data frame allowing up to 256 events to be embedded into the Timing stream. This was originally done with a mix of analog and digital circuitry that created a 'bang-bang' transceiver to recover both a phase stable 10MHz clock, and 8b TCLK events. In later years the circuitry was embedded in a custom Fermilab Integrated Circuit (IC), thousands of which are scattered across the complex today.

In this scheme, the 10 MHz carrier signal is phase-modulated: every 100 ns bit cell has a transition at the cell boundary, and a mid-cell transition encodes a data "1" (no mid-cell transition for "0"). Each TCLK Event word is 10 bits long: it begins with a start bit (always 0), followed by an 8-bit Event code, and ends with a parity bit. The transmitter enforces a minimum gap of two 1 bits between events. If no events are being sent, TCLK defaults to an idle pattern (continuous 1s) which appears as a 10 MHz square wave on the line. This guarantees that devices can always lock onto the carrier. In practice, events cannot occur closer than about 1.2 µs apart due to the frame length and required spacing

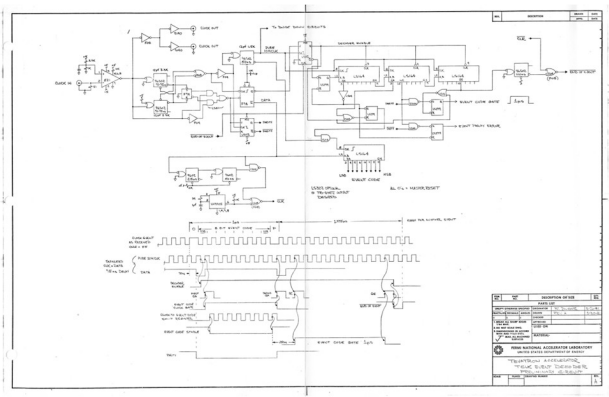

*Figure 3: Original TCLK Decoder Schematic*

Somewhat unique to TCLK's implementation is the ability to be modulated by the accelerator's RF. For the 53MHz Main Injector, the 7th sub-harmonic of 7.5MHz is used as the carrier frequency, allowing receivers to apply the same hardware to decode events synchronously with the beam. This encoding is periodically re-aligned to the local RF at the various service buildings, providing enough stability to facilitate injection and extraction at any point along the machine. The TCLK implementation for Main Injector is referred to as 'Main Injector Beam Sync (Clock)' or MIBS, the same for Recycler (RRBS).

## MAJOR SYSTEMS

There are five major systems that interact with Fermilab's Timing infrastructure. These are listed here:

**The Timeline Generator** – This system calculates and loads the series of Events, or "Timeline", to be played out during subsequent machine cycles. The Timeline Generator (TLG) acts as the user application for the Timing system to provide a readily configurable interface to the underlying hardware. This system will be upgraded with the PIP-II transition to EPICS.

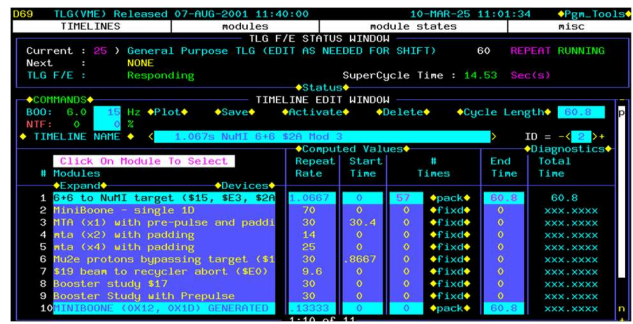

*Figure 4: TLG User Interface*

**ComEd** – This is the public utility that delivers electricity to Fermilab's campus. Fermilab is fed by two substations referred to as 'Master Sub' and 'Kautz Road' for Booster and Main Injector respectively. All accelerator systems at Fermilab are subject to the ambient conditions of the public utility, and drift with the 60Hz power over the course of the day.

**GMPS** - The Fermilab Booster is *highly* dependent on ComEd line frequencies to optimize ramp times with the 15Hz subharmonic (future 20Hz). Booster regulates its bend bus through the Gradient Magnet Power Supply (GMPS) system where a measurement of the local magnetic field (B-Field) triggers injection at minimum intensity, leveraging resonant mains power to control gradient magnets in an efficient manner [4]. Booster supplies do not have sufficient power to regulate these magnets otherwise. The introduction of 'flat injection' in the PIP-II era will further complicate this regulation challenge.

**Cogging** – Cogging at Fermilab refers to the process of synchronizing the relative phase (position) of proton bunches between Booster and Main Injector/ Recycler. The Cogging Module adjusts the Timing of RF buckets so that the extracted beam from one ring arrives in alignment with empty buckets in the receiving ring, avoiding losses and ensuring efficient injection. This is achieved by varying the RF frequency of Booster in conjunction with the GMPS sweep. Beam-synchronous clocks like MIBS and RRBS, along with Timing events triggered over TCLK, coordinate cogging operations, allowing for turn-by-turn precision and automated control during high-intensity operations.

**MECAR** – Main Injector Extraction Control and Regulation (MECAR) and its communication protocol 'MDAT' (Machine Data) are two critical systems at Fermilab that complement the Event-driven TCLK system by broadcasting real-time machine parameters around the ring. MECAR's predecessor 'TECAR' (Tevatron-CAR) was originally used to share analog reference waveforms to the various corrector controllers regulating the TeV ramp. MECAR expanded on this by broadcasting beam energy and momentum, enabling these corrector controllers to generate higher order waveforms on top of the curve.

**General Order of Operations** – The systems outlined above come together in the following manner.

1. BMIN from GMPS triggers the Booster Reset, starting the chain of events.
2. The Beam Switch Sum Box (BSSB) validates permits are in place and starts beam.
3. Booster captures beam from Linac and accelerates with the GMPS curve.
4. At near-peak momentum, Booster cogs to Main Injector's RF system for transfer.
5. Once cogged, Booster extracts to MI through the BES (Beam Extraction Sync) signal.
6. This repeats at 15Hz for however many Booster batches are to be ramped.
7. At a set time, TCLK initiates the MI ramp, which is re-timed to MIBS/MECAR.
8. Main Injector accelerates the beam from 8GeV to 120GeV along a preconfigured curve.
9. Beam is extracted from the accelerator.

## PRESENT

Today TCLK is decoded in real time by the Field Programmable Gate Array (FPGA) hardware deployed in Fermilab's accelerator systems since the 1990's. This approach provides a far more flexible and integrated method of observing accelerator events. Of note however is the impact of decoding by oversampling through the FPGA's local clock. Without the 'Bang-Bang' analog circuitry, jitter is imparted in the 10MHz edge proportional to the FPGA clock period (4~10ns). While negligible for most systems, the original Fermilab IC can still be considered the 'Gold Standard' of TCLK decoding.

This minute technological factor is in part the reason behind continued use of Tawzer modules (c. 1987) for Booster extraction [5]. Despite TCLK's exceptional accuracy and the high level of precision provided by newer FPGA's, Fermilab's Booster kickers still rely on a dedicated RF synchronization functionality for efficient operation.

## EVOLUTION (XCLK)

With the advent of Project X (Now PIP-II) in the early 2000's, a new 'XCLK' Timing system was proposed. This system would use high speed transceivers to both increase resolution (more events per second), and allow for encoding of data within the Event stream. The implementation was otherwise the same, with Manchester encoding over a 1Gbps link providing 500Mbps of throughput tightly synchronized to the current TCLK timeline. This protocol would form the basis of the upgraded 'Accelerator Clock' (ACLK).

## ACCELERATOR CLOCK (ACLK)

Designed in conjunction with the PIP-II project, ACLK is the successor to TCLK. This modernized Timing system encompasses all of TCLK's legacy functionality, in addition to provisioning for future operating scenarios. Every effort has been made to bring Fermilab's Timing system in line with industry standards while facilitating a seamless transition for the complex into the 21$^{st}$ century.

ACLK utilizes the same serial Event distribution stream and star topology as its predecessor, allowing for events to be injected in real time to the Timing system with an optimal, bounded latency through the network.

To facilitate full backwards compatibility, ACLK also utilizes a similar 10MHz base frequency, but in this instance the frequency represents the spacing between Event Windows (100ns) as opposed to the frequency at which the data is encoded. The end result yields legacy events that can be decoded from ACLK to TCLK with similar precision to the original system.

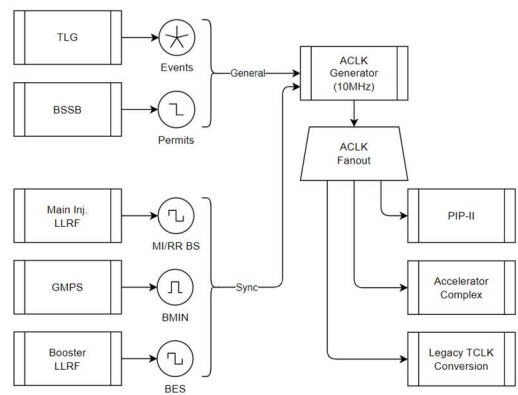

*Figure 5: ACLK System Diagram*

**Frame Structure** – The ACLK protocol utilizes 96b frames in the format shown below:

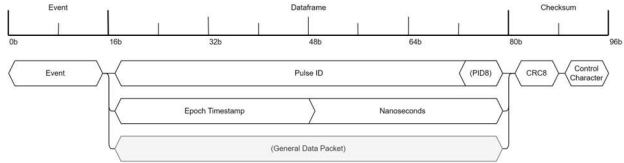

*Figure 6: ACLK Frame Structure*

This structure contains the following information:
- 16b Event – Machine reset, abort, extraction, etc.
- 64b Data packet – Timestamp, beam intensity, etc.
- 8b CRC – Cyclic Redundancy Check
- 8b Control Character – Marks frame boundary

## ACLK VS. TCLK

To summarize: bandwidth is the primary delineation between TCLK and its successor. Operating at more than 100x the speed of TCLK, ACLK enables users to encode both events and data frames within the span of a single TCLK transition. This improves the accuracy of legacy events encoded into the stream, and opens the door to higher repetition rates without the risk of serial congestion.

## ACLK VS. LCLK

Linac Clock (LCLK, technically LCLK-II) is the Beam Synchronous implementation of ACLK for PIP-II, similar to the relation between TCLK and MIBS. LCLK's transmission frequency is derived from the 162.5MHz Master Oscillator (MCO) of PIP-II, with Event windows at the 16$^{th}$ sub-harmonic, or 10.16MHz. Coupling the clock frequency to the RF in this manner allows for BS Events to arrive reliably inter-bucket without crossing clock domains. Additionally, selection of a frequency near the ACLK 10MHz reference allows for a common FPGA core to be utilized between machines. Apart from the difference in frequency, LCLK is functionally the same as ACLK.

## WHITE RABBIT INTEGRATION

To readily address the intrinsic compound errors with real-time Timing distribution, local synchronization via White Rabbit *may* be utilized at every node. This industry standard protocol re-aligns the 10MHz carrier frequency and provide an absolute time reference for nearby devices. In this scenario events are distributed in real time via ACLK, and the White Rabbit IP network provides synchronization and connectivity as shown below. This provides essential Internet Protocol (IP) communication capabilities for configuration of the Timing units, as well as an established off-the-shelf solution for time distribution over network infrastructure. Note: WR links are not required to maintain the precision of short-run Timing distribution between nodes and devices.

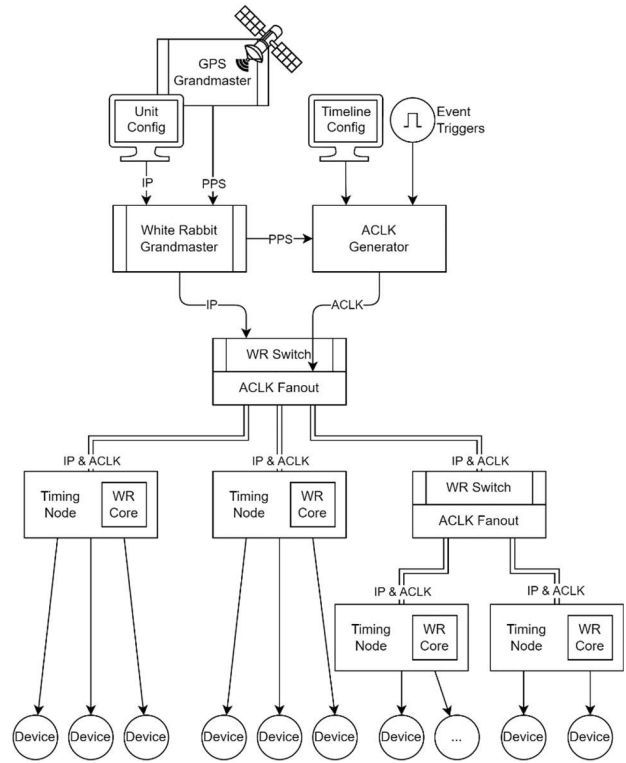

*Figure 7: White Rabbit & ACLK Distribution*

## A LOOK AT REAL TIME

The need for real time control was acknowledged early in TCLK's design with development of the accompanying Timeline Generator (TLG). The TLG configures a queue of TCLK events to play out on a cycle-by-cycle basis. The original software release [6] speaks to their decision to operate the facility in a state based manner, rather than a time based manner as CERN's Proton Synchrotron (PS) machine was implementing at the time:

> *[…regarding deterministic timelines at Fermilab] We have decided to approach the problem in a different way for two reasons. The first is that it would demand an unacceptable expansion of the control system, both in hardware and software. The second comes from detailed investigation of those parameters which need to change for different types of cycles in the MR and Booster. In fact, we believe the number of parameters and function tables which have to be swapped is not large and each parameter should be considered individually for the best method.*

Four decades later, this analysis was revisited for the Accelerator Operations and Control Research Network (ACORN) project. In conjunction with the active Low Level RF (LLRF) upgrades for the Booster Accelerator RF (BARF) system [7], it was determined that the existing high power regulation infrastructure would not be capable of operating in this manner without complete overhaul, and 'individual consideration' of parameters on a cycle-by-cycle basis still provides the optimal approach to machine Timing for the complex.

## STATUS

The ACLK, LCLK, and White Rabbit components outlined above are to be implemented and validated under the scope of Fermilab's PIP-II project. PIP-II will provide the engineering support to transition Fermilab's Timing infrastructure from TCLK to ACLK; distribution of the protocol beyond PIP-II will be the responsibility of Fermilab's control system modernization project, ACORN. The two projects have worked extensively to deliver a common upgrade path for Fermilab's Timing infrastructure.

## FUTURE

With a fully synchronized Timing system controlling Fermilab's campus, orchestration of the complex can be addressed in a more deterministic manner. While it is impossible to know in advanced the optimal Timing of a single Booster injection, correlation and predictive methods can be utilized to fine tune operation of the machine over time.

Consider the simplest scenario: one Booster cycle. With perfect repeatability, the RF and power programs can be programmed independently as the magnets will reach optimal bend strength at predictable 20ns bunch intervals. This 20ns precision immediately devolves however as power is utilized in differing capacities upstream by Main Injector or other public utility users. These interactions impact the power grid on timescales from seconds (one Main Injector ramp), to hours (the "Warrenville Toaster Effect"), to years (2024's Kautz Road Outage), Fortunately, the impact is quickly reflected in the succeeding timeline by the various open and closed loop systems that keep Fermilab synchronized.

The order of Timing Events broadcast during the preceding cycle is captured concisely in the 'Manifest', a Multicast internet packet containing the exact series of Event / Timestamp pairs. By logging the manifest over successive runs, researchers can assess performance of the machine and measure the interactions between major accelerator systems relative to the start of cycle. Application of statistical methods or *Machine Learning* enables prediction of successive timelines from discrete measurements. This has two notable applications outlined here.

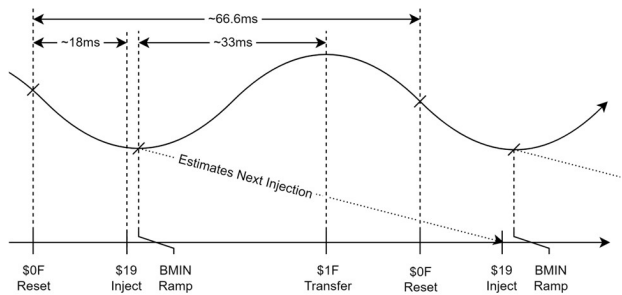

Figure 8: Approximate Timing of a Booster Cycle

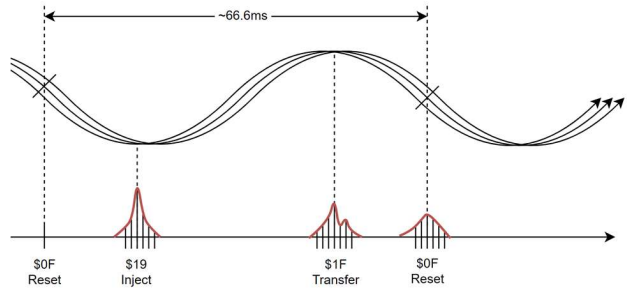

Figure 9: Statistical Aggregation of Booster Events

## FAR-SITE SYNCHRONIZATION

A fundamental limitation to real-time correlation of accelerator events between the near & far detectors is the propagation delay incurred by electrical signals relative the arrival time of the neutrino beam. This necessitates the use of large acquisition windows, deep buffers, and post-processing to filter out interactions not tied to the beam arrival window. A Digital Twin of Fermilab's timeline, operating on the timestamps of preceding events over numerous machine cycles has the potential to play out remotely, providing high precision gating windows with sub-microsecond accuracy. While this approach does not have the required dynamicism for local regulation, it is capable of providing reliable triggers through faster than light prediction.

## EVENT VECTORIZATION

Cycle-by-cycle assessment of Fermilab's manifest has the added benefit of identifying trends in Event Timing that may otherwise be opaque to users & the TLG. A Digital Twin of the Timing system in conjunction with ACLK's enhanced data packet capabilities allows clock generation hardware to broadcast the general direction, or vector the Event is moving so that downstream systems may make precision adjustments accordingly. This is exceptionally useful in transfer scenarios where kicker systems have a finite window for injection and extraction. Vector encoding within the Event itself enables system-by-system tuning according to the Accelerators larger trends.

## CONCLUSION

TCLK has had an indelible impact in the operation of Fermilab. The underlying mechanisms of the Timing system have enabled reliable operation for more than 40 years, and present a significant modernization challenge to the Controls group. Strategic design decisions have yielded a Timing system that is both forwards & backwards compatible, enabling advanced regulation capabilities. All stemming from choices made before the Tevatron was commissioned. Please consider anthropologist Marshall McLuhan:

*We shape our tools, and thereafter our tools shape us.*
*McLuhan, 1967*